\documentclass[%
reprint,
superscriptaddress,
amsmath,amssymb,
prd,
]{revtex4-1}

\usepackage{graphicx,wrapfig,lipsum}
\usepackage{subeqnarray}
\usepackage{url,hyperref} 
\usepackage{color}
\usepackage{ulem}
\usepackage{verbatim}

\usepackage{natbib}
\newcommand{\req}[1]{Eq.\,({\ref{#1}})}
\newcommand{\reqs}[2]{Eqs.\,({\ref{#1}},{\ref{#2}})}

\begin{document}
\title{Radiation reaction and limiting acceleration}
\author{Will Price}
\email{wprice@email.arizona.edu}
\affiliation{Department of Physics, The University of Arizona, Tucson, AZ, 85721, USA}
\author{Martin Formanek}
\email{martinformanek@email.arizona.edu}
\affiliation{Department of Physics, The University of Arizona, Tucson, AZ, 85721, USA}
\author{Johann Rafelski}
\email{johannr@email.arizona.edu}
\affiliation{Department of Physics, The University of Arizona, Tucson, AZ, 85721, USA}

\date{January 13, 2022}
%
\begin{abstract}
We investigate the strong acceleration properties of the radiation reaction force and identify a new and promising limiting acceleration feature in the Eliezer-Ford-O'Connell model; in the strong field regime, for many field configurations, we find an upper limit to acceleration resulting in a bound to the rate of radiation emission. If this model applies, strongly accelerated particles are losing energy at a much slower pace than predicted by the usual radiation reaction benchmark, the Landau-Lifshitz equation, which certainly cannot be used in this regime. We explore examples involving various ``constant'' electromagnetic field configurations and study particle motion in a light plane wave as well as in a material medium.
%
%
%
\end{abstract}
\maketitle

\section{Introduction}
Inspired by the Born-Infeld (BI) theory of electromagnetism~\cite{Born:1934gh,Rafelski:1973fm,Bialynicki-Birula:1983}, we ask if there can be a natural upper limit to the acceleration that a charged particle can experience. In the BI model, this bound can be achieved by introducing an upper limit to the electromagnetic field strength. For particle motion in certain field configurations, the upper bound on field strength leads to an upper bound on the acceleration. We identify the presence of an upper bound to acceleration in the Eliezer-Ford-O'Connell (EFO) description of radiation reaction (RR) (see, for example,~\cite{Kravets:2013,Kravets:2014lca,Burton:2014wsa}) for nonhyperbolic motion. We study the resulting (classical) charged particle dynamics for a few simply soluble force field configurations: (a) constant fields, (b) plane waves, and (c) a covariant material friction force. We compare results to another well-known description of the RR force, the Landau-Lifshitz (LL) equation~\cite{LL:1962}.

An accelerated (relativistic) charge emits radiation and loses energy at a rate given by the relativistic Larmor formula~\cite{Jackson:1998nia}
\begin{equation}
 P = -m\tau_0 a_\mu a^\mu \, ,\qquad \tau_0 = \frac{2}{3}\frac{e^2}{4\pi\epsilon_0 mc^3} =6.27 \times 10^{-24}\,\mathrm{s}\,.
\end{equation}
Here and below, the numerical values are obtained using the mass and charge of an electron. This means that a charged particle experiencing an acceleration of the magnitude
\begin{equation}\label{aclas}
 a_{\text{RR}} = \frac{c}{\tau_0}=4.78\times 10^{31}\,\mathrm{\frac{m}{s^2}} \, 
\end{equation}
emits an energy equal to its rest mass equivalent in a span of time equal to the characteristic RR time interval $\tau_0$
\begin{equation} \label{eq:plim}
P_{\text{RR}} = \frac{mc^2}{\tau_0} \, .
\end{equation}
This situation has motivated our search for a RR force with properties akin to the BI theory; if a particle's acceleration was bounded by $a_{\text{RR}}$,~\req{aclas}, the particle's radiation rate is bounded by \req{eq:plim}. For such bounded radiation emission~\req{eq:plim} the EFO equation generates a RR force that in turn generates a limited acceleration~$a_{\text{RR}}$, creating a physically self-consistent model of charged particle motion.

The conventions used in this paper are as follows. We use a flat spacetime metric
\begin{equation}
 g_{\mu\nu} = \text{diag}(1,-1,-1,-1) \, .
\end{equation}
The external field is given by the electromagnetic (EM) tensor
\begin{equation}\label{eq:EMtensor}
F^{\mu\nu} \equiv \partial^\mu A^\nu - \partial^\nu A^\mu = \left(\begin{matrix}
0 &-\pmb{\mathcal{E}}/c\\
\pmb{\mathcal{E}}/c & -\epsilon_{ijk}\mathcal{B}^k
\end{matrix} \right)\,,
\end{equation}
where $\epsilon_{123} = +1$ and the 4-potential is given by $A^\mu = (V/c,\pmb{A})$, where $V$ is the scalar potential and $\pmb{A}$ is the vector potential. $\pmb{\mathcal{E}}$ and $\pmb{\mathcal{B}}$ are the electric and magnetic field vectors which can be computed from the 4-potential components as
\begin{equation}
\pmb{\mathcal{E}} = - \nabla V - \frac{\partial \pmb{A}}{\partial t}\,, \quad \pmb{\mathcal{B}} = \nabla \times \pmb{A}\,.
\end{equation}
The dual EM tensor is defined by
\begin{equation}
\widetilde{F}_{\alpha\beta} \equiv \frac{1}{2}\epsilon_{\alpha\beta\mu\nu}F^{\mu\nu},
\end{equation} 
with the totally antisymmetric symbol defined by $\epsilon_{0123} = +1$.

For constant fields, we will use the following electric and magnetic constants with units of frequency:
\begin{equation}
\Omega_E = \frac{e\mathcal{E}}{mc}\,, \quad \Omega_B = \frac{e\mathcal{B}}{m} \,,
\end{equation}
where $e$ and $m$ are the charge and mass of the particle, respectively.

We also define the following EM field invariants $\mathcal{S}$ and $\mathcal{P}$ 
\begin{align}\label{eq:S}
\mathcal{S} &\equiv \frac{1}{4}F^{\mu\nu} F_{\mu\nu} = \frac{1}{2}(\mathcal{B}^2 - \mathcal{E}^2/c^2)\,,\\
\label{eq:P}\mathcal{P} &\equiv \frac{1}{4}F^{\mu\nu} \widetilde{F}_{\mu\nu} = \pmb{\mathcal{B}}\cdot \pmb{\mathcal{E}}/c\,.
\end{align}

In Sec.~\ref{Sec:EFO}, we reintroduce the Eliezer-Ford-O'Connell RR force. A comparison of key features of different models of the RR force is presented in Sec.~\ref{Sec:CompRR}. For spacetime-independent field configurations we present in Sec.~\ref{sec:const_fields} a general result determining the form of the acceleration magnitude in terms of the invariant Lorentz force acceleration and field invariants; we then describe special dynamical examples in different field configurations. Among nonconstant fields, the configuration of particular physical relevance is the plane electromagnetic wave explored in Sec.~\ref{sec:lightM}. Another case of interest is a friction force due to passage through matter~\cite{Formanek:2020zwc} which is considered in section~\ref{sec:material_friction}. In section~\ref{sec:quantum}, we discuss the connection to the limiting EM field strength recognized in quantum electrodynamics. We summarize our results and provide future outlook in Sec.~\ref{Sec:Final}. 

\section{Eliezer-Ford-O'Connell RR Force}\label{Sec:EFO}
In this section we present the EFO form of the radiation reaction force. For a particle subject to an external force $\mathcal{F}_\text{ext}^\mu$, the EFO equation takes the following form:
\begin{equation} \label{eq:FO}
 ma^\mu = \mathcal{F}_\text{ext}^\mu + \tau_0 P^\mu_\nu \frac{d}{d\tau}\mathcal{F}_\text{ext}^\mu \, ,
\end{equation}
where the projection tensor is given by 
\begin{equation}
 P^\mu_\nu = \delta^\mu_\nu - \frac{u^\mu u_\nu}{c^2} \, ,
\end{equation}
which projects onto the hypersurface orthogonal to the 4-velocity $u^\mu$. We will show that, outside of a few specific external field configurations,~\req{eq:FO} generally predicts that particles cannot reach an acceleration greater than the limiting value $a_\mathrm{RR}$. 

The EFO equation is known to correspond with the LL RR force~\cite{LL:1962} in the leading perturbative limit (for the most recent studies see \cite{Kravets:2013, Kravets:2014lca, Garcia:2015}), and we show this explicitly through several examples. This demonstrates that the EFO model can be thought of as a strong field, limiting acceleration generalization of the LL model. However, there are exceptional cases that do not conform to the limiting acceleration principle which are of special interest. 

The EFO equation of motion was first studied by Eliezer in 1948~\cite{Eliezer:1948} and later by Ford and O'Connell in 1991~\cite{Ford:1991, Ford:1993}. Both studies derived the EFO equation as an approximate RR equation of motion for a charged particle with an extended size. To the best of our knowledge, the important consequence of this equation, the upper limit on the acceleration, is for the first time discussed here. It also should be noted that in the absence of Google Scholar, we would have rediscovered the EFO-RR force for a third time, as we were constructing a RR force based on the desired properties: a hybrid format involving fields and acceleration leading to a limiting acceleration. These are uniquely satisfied by the EFO form.

For a particle in an EM field with the external force given by the Lorentz force
\begin{equation}
 \mathcal{F}_\text{ext}^\mu = eF^{\mu\nu}u_\nu \, ,
\end{equation}
the EFO equation takes the form
\begin{equation} \label{eq:FO2}
 ma^\mu = eF^{\mu\nu}u_\nu + e\tau_0\left(u^\alpha \partial_\alpha F^{\mu\nu}u_\nu +P^\mu_\nu F^{\nu\alpha}a_\alpha\right) \, .
\end{equation}
Grouping the acceleration-dependent terms and defining the tensor
\begin{equation}
 M^\mu_\nu \equiv \delta^\mu_\nu - \frac{e\tau_0}{m}P^{\mu\alpha}F_{\alpha\nu} \, ,
\end{equation} 
we obtain the following expression:
\begin{equation} \label{eq:FO3}
 M^\mu_\nu a^\nu = \frac{e}{m}\left(F^{\mu\nu}+\tau_0 \frac{d}{d\tau}F^{\mu\nu}\right)u_\nu \, .
\end{equation}
In order to integrate~\req{eq:FO3} and calculate the motion of a particle, the tensor $M^\mu_\nu$ must be invertible to allow us to solve for the acceleration 4-vector $a^\mu$, orthogonal to 4-velocity $a\cdot u = 0$. Since the right side of~\req{eq:FO3} is also orthogonal to the 4-velocity $u^\mu$, the inverse $(M^\mu_\nu)^{-1}$ only needs to act on the subspace orthogonal to $u^\mu$. Due to this fact, an inverse can be found and is done so in Ref.~\cite{Kravets:2014lca}. This allows one to rewrite~\req{eq:FO3} with the 4-acceleration expressed purely in terms of the EM field and the 4-velocity.

\section{Comparison of RR models}\label{Sec:CompRR}
\noindent
Before exploring the predictions of the EFO equation, we discuss it in the context of the standard formulation of radiation reaction for weak acceleration. 
\begin{itemize}
\item
The covariant form of Larmor radiation power loss cast into reaction force form is the Lorentz-Abraham-Dirac (LAD) equation~\cite{Dirac:1938nz}
\begin{equation} \label{eq:LAD}
 ma^\mu = eF^{\mu\nu}u_\nu + m\tau_0 \left( \frac{da^\mu}{d\tau}+\frac{a^2}{c^2}u^\mu \right) \, .
\end{equation}
To obtain LAD, the self-action of the radiation field of a charge on itself needs to be allowed for~\cite{Teitelboim:1971gb}. This procedure is questionable for point particles and there is a hefty price to be paid for this: \req{eq:LAD} is known to allow for runaway solutions, {\it i.e.\/} motion in which particles accelerate exponentially irrespective of the applied force. Since LAD includes derivatives of acceleration, causality is not guaranteed and indeed is violated in general~\cite{Rohrlich:1965-90}. The LAD equation of motion, when considered as an initial value problem, requires an additional boundary condition aside from the initial position and momentum.
\item
For all the above reasons, in most applications of RR, one encounters the form presented originally in the LL textbook~\cite{LL:1962} as a reduction of order approximation of the LAD equation by making repeatedly the substitution $a^\mu \rightarrow \frac{e}{m}F^{\mu\nu}u_\nu$ on the right-hand side of~\req{eq:LAD}
\begin{align} \label{eq:LL}
 ma^\mu = eF^{\mu\nu}u_\nu &+ e\tau_0 \Big(u^\alpha \partial_\alpha F^{\mu\nu}u_\nu \\
 \nonumber &+ \frac{e}{m}P^{\mu\nu}F_{\nu\alpha}F^{\alpha\beta}u_\beta \Big) \, .
\end{align}
The LL equation can be also derived as a first order RR correction to the Lorentz force as shown in~\cite{Gralla:2009}. This method does not rely on the LAD equation as a starting point and therefore establishes the LL equation as a trustworthy RR force for weak acceleration independent of the LAD model. 
\end{itemize}

In some translated editions of the Landau and Lifshitz textbook, these authors appeared as claiming that the LAD and LL formulations were equivalent. Clearly, this is not the case in general, as the LAD equation contains unphysical solutions absent from the LL equation \cite{Spohn:2004ik}. Unlike the LAD formulation, which aims to be a fully consistent RR force, the LL model is a manifestly perturbative description of RR applicable only in the weak acceleration domain.

We are however interested in studying the strong acceleration domain, beyond the applicability of the LL equation. To formulate an equation capable of this, as a first step, we note that the LAD equation can be equivalently written as 
\begin{equation}
 ma^\mu = eF^{\mu\nu}u_\nu + \tau_0P^\mu_\nu\frac{d}{d\tau}(ma^\nu) \, . 
\end{equation}
If we then make the substitution $ma^\nu \rightarrow eF^{\nu\alpha}u_\alpha$, we arrive at 
\begin{equation}\label{eq:EFObis}
 ma^\mu = eF^{\mu\nu}u_\nu + e\tau_0P^\mu_\nu\frac{d}{d\tau}(F^{\nu\alpha}u_\alpha) \, ,
\end{equation}
which is the EFO equation. 

We recall that the LL equation is obtained from LAD as a reduction of order approximation~\cite{LL:1962} obtained by replacing the acceleration with the applied force. Our argument seen in~\req{eq:EFObis} could be interpreted in the same way; the EFO equation follows using such a (modified) procedure. However, the EFO equation was originally derived through a different line of thinking: as a RR equation of motion for extended particles. As long as we lack a satisfactory first principles treatment of RR for point particles we cannot view any RR version as being more fundamental than another. However, EFO appears to be more justified by first principles as well as more elegant in its form. Therefore, for the purposes of this work, we will assume the EFO equation as being the best approximate to the exact, yet to be discovered, RR particle dynamics formulation.

Comparing the EFO equation to the LAD and LL equations, we emphasize the hybrid EFO format: LAD places RR alone as an effect of dynamic motion using 4-velocity and 4-acceleration to describe RR; on the other hand, LL only uses EM fields to characterize the RR. The EFO format employs all available 4-vectors $u^\mu, a^\mu, \mathcal{F}^\mu_\text{ext}$, including both dynamical variables and fields, and thus creates a ``hybrid'' formulation conceptually lying between the LAD and LL formalism.

It is easy to see that the EFO equation leads to the LL equation in a perturbative iteration using $\tau_0$ as a smallness parameter. We will therefore assume the EFO equation as the full RR force, and, in light of the equivalence between the LL and EFO equations for weak acceleration, we will view the LL equation as a perturbative approximation of the EFO equation. The hierarchy of these equations can be seen in Table~\ref{tab:RR_models}. 

\begin{table*} 
\begin{tabular}{c|ccc}
 \hline
Name & Covariant equation & Year & Reference\\
 \hline \hline \noalign{\vskip 0.1cm}
Lorentz-Abraham-Dirac (LAD) & $m a^\mu = \mathcal{F}^\mu + \tau_0 P^\mu_\nu \displaystyle\frac{d}{d\tau}(ma^\nu)$ & 1938 &~\cite{Dirac:1938nz}\\
Eliezer-Ford-O'Connell (EFO) & $m a^\mu = \mathcal{F}^\mu + \tau_0 P^\mu_\nu \displaystyle\frac{d}{d\tau}(eF^{\nu\alpha}u_\alpha)$ & 1948, 1991 &~\cite{Eliezer:1948},~\cite{Ford:1991},~\cite{Ford:1993}\\[0.2cm]
 \hline	\noalign{\vskip 0.1cm}		 
Landau-Lifshitz (LL) & $ma^\mu = \mathcal{F}^\mu + \tau_0\left( e\displaystyle\frac{d}{d\tau}(F^{\mu\nu})u_\nu + \frac{e^2}{m} P^\mu_\nu F^{\nu\alpha} F_{\alpha\beta} u^\beta\right)$ & 1962 &~\cite{LL:1962}\\
 Mo-Papas (MP) & $ma^\mu = \mathcal{F}^\mu + e\tau_0 P^\mu_\nu F^{\nu\alpha} a_\alpha$ & 1971 &~\cite{TseChin:1971tmj}\\[0.2cm]
 \hline 
\end{tabular}
 \caption{\label{tab:RR_models} Classical radiation reaction models for electromagnetic force $\mathcal{F}^\mu = e F^{\mu\nu}u_\nu$. Top: the LAD and EFO forms which attempt complete description. Bottom: EFO approximations: LL for `small' $\tau_0$; Mo-Papas for constant fields.}
\end{table*}

The last entry in the table, the Mo-Papas equation \cite{TseChin:1971tmj}, is the constant field form of the EFO form. It cannot be viewed as another potentially valid RR force, as it is missing a field derivative term and thus disagrees with LL equations for weak fields, leading to unphysical solutions. It has been shown~\cite{Huschilt:1974} that the Mo-Papas radiation reaction force vanishes for any motion in one dimension, an unphysical result not seen in other RR formulations.

\section{Homogeneous and Constant Fields}\label{sec:const_fields}
\subsection{Acceleration as a function of the Lorentz force and field invariants}
\noindent
For constant, homogeneous EM fields, the $u\cdot \partial F^{\mu\nu}u_\nu = u_\nu dF^{\mu\nu}/d\tau$ term in EFO~\req{eq:FO} and LL~\req{eq:LL} vanishes. The EFO equation can then be written as 
\begin{equation}\label{eq:MoPapas}
a^\mu = \frac{e}{m}F^{\mu\nu}u_\nu + \tau_0 \frac{e}{m} P^\mu_\nu F^{\nu\alpha}a_\alpha \,;
\end{equation}
{\it i.e.\/} the Eliezer-Ford-O'Connell equation reduces to the Mo-Papas equation; compare with Table~\ref{tab:RR_models}. Therefore, all our analysis in this section, including the result of limiting acceleration, will also apply to the Mo-Papas equation. 

We will now show that the Eliezer-Ford-O'Connell equation of motion can be explicitly solved for the invariant square of the 4-acceleration for constant fields. First, let us denote the invariant quantities
\begin{align}
C_1 \equiv uFFa\,, &\quad C_2 \equiv aFFa\,,\\
C_3 \equiv uFa\,, &\quad C_4 \equiv u\widetilde{F}a\,,
\end{align}
where in our notation $uFFa=u_\mu F^{\mu\nu}F_{\nu\alpha}a^\alpha$, and so on. We can then multiply~\req{eq:MoPapas} with the 4-vectors $a_\mu$, $(uF)_\mu$, $(u\widetilde{F})_\mu$, and $(uFF)_\mu$, to derive the following linear equations in terms of the introduced invariants:
\begin{align}
\label{eq:sysfirst}a_\mu: &\quad a^2 = - \frac{e}{m}C_3\,,\\
(uF)_\mu: &\quad C_3 = \frac{e}{m}uFFu + \frac{e\tau_0}{m}C_1\,,\\
\label{eq:systhird}(u\widetilde{F})_\mu: &\quad C_4 = - \frac{e}{m}\mathcal{P}c^2\,,\\
\label{eq:syslast}(uFF)_\mu: &\quad C_1 = \frac{e\tau_0}{m}\left(-\mathcal{P}C_4 - 2 \mathcal{S}C_3 - \frac{C_3}{c^2}uFFu \right)\,.
\end{align}
The identities
\begin{align}
(F\widetilde{F})^\mu_\nu &= (\widetilde{F}F)^\mu_\nu = - \mathcal{P} \delta^\mu_\nu\,,\\
(F^3)^\mu_\nu &= - \mathcal{P}\widetilde{F}^\mu_\nu - 2 \mathcal{S}F^\mu_\nu
\end{align}
were used to simplify the expressions. 
 
From the system of equations~\req{eq:sysfirst}-\req{eq:syslast}, we can eliminate the invariants $C_1$, $C_2$, $C_3$, and $C_4$ by substituting from~\req{eq:sysfirst}-\req{eq:systhird} to~\req{eq:syslast}. We then define the Lorentz force acceleration as 
\begin{equation}
 a^\mu_{LF} = \frac{e}{m}F^{\mu\nu}u_\nu \, ,
\end{equation}
so that we can express $a^2$ as a function of $a_{LF}^2$,\, $\mathcal{S}$, and $\mathcal{P}$, yielding 
\begin{equation}\label{eq:a2resummed}
 a_\mu a^\mu = a^2_{LF} \displaystyle\frac{1+\tau_0^2\frac{e^4}{m^4}\displaystyle\frac{c^2\mathcal{P}^2}{|a^2_{LF}|}}{1+\tau_0^2(\frac{e^2}{m^2}2\mathcal{S}+\frac{|a^2_{LF}|}{c^2})} \, .
\end{equation}
By inspecting~\req{eq:a2resummed}, we can see that the presence of limiting acceleration is dependent on the behavior of the two field invariants ($\mathcal{S}$ and $\mathcal{P}$) and the dynamical invariant
\begin{equation}\label{eq:a2lf}
 a^2_\mathrm{LF} = - \frac{e^2}{m^2}uFFu\,.
\end{equation}
The Lorentz force acceleration is a spaceike vector, so $a^2_\mathrm{LF} < 0$ at all times. 

For comparison, the square of the acceleration for the LL equation can be written as 
\begin{equation} \label{eq:a2_LL}
 a_\mu a^\mu = a_{LF}^2 \left(1-\tau_0^2\left(\frac{e^4}{m^4}\frac{c^2\mathcal{P}^2}{|a_{LF}^2|}-2\frac{e^2}{m^2}\mathcal{S}-\frac{|a_{LF}^2|}{c^2}\right)\right) \, .
\end{equation}
The LL acceleration does not have an upper bound in strong fields. When compared to the EFO result~\req{eq:a2resummed}, the sign of the lowest order correction (quadratic in $\tau_0$) differs. This supports our assumption that the EFO equation lowers the invariant magnitude of acceleration compared to the Lorentz force case whereas the LL equation raises it, in certain field configurations.

In the rest of this section we will discuss examples characterized to a large extent by specific values of $\mathcal{S}$ and $\mathcal{P}$ in order to understand exactly under what circumstances limiting acceleration can arise.

\subsection{Motion parallel to a constant electric field}\label{sec:EConst}
For a pure electric field and motion in one dimension (1D), $\mathcal{P} = 0$ and $\mathcal{S}<0$. Moreover, by squaring the Lorentz force we also find in the current case
\begin{equation} \label{eq:a2_1d}
\frac{e^2}{m^2}\,2\,\mathcal{S}=- \frac{|a^2_{LF}|}{c^2} \,. 
\end{equation}
By~\req{eq:a2resummed} this implies
\begin{equation} \label{eq:a2_1d_2}
a_\mu a^\mu = a^2_{LF}\,.
\end{equation}
There is then no EFO radiation reaction for a charge uniformly accelerated along electric field lines, which is also a well known feature of the Landau-Lifshitz RR model. Motion in the case of either model is then governed solely by the 1D Lorentz force and the acceleration is unbounded (in the literature referred to as hyperbolic motion). 

The question of radiation reaction in a 1D constant electric field has a long history~\cite{Fulton:1960} as a well-known controversy of radiation reaction models and the problem remains unsolved by the EFO equation. We will continue this discussion in concluding Sec.~\ref{Sec:Final}.

\subsection{Constant magnetic field ($\mathcal{S}>0,\, \mathcal{P}=0$)}\label{sec:Bconst}
Let us consider a constant magnetic field in the $z$-direction $\pmb{\mathcal{B}} = (0,0,\mathcal{B})$. For a particle with velocity in the plane perpendicular to the field we have
\begin{align}
u^\mu &= (\gamma c, u_x, u_y, 0) = \gamma c(1, \beta_x, \beta_y, 0) \, , \\ 
a^\mu &= (\dot{\gamma}c,\dot{u}_x,\dot{u}_y,0) \, ,
\end{align}
in which the dot here refers to the proper time derivative. Rewriting the Eliezer-Ford-O'Connell equation~\req{eq:FO} with the acceleration terms on the left side, we obtain 
\begin{equation}\label{eq:FO_B}
 a^\mu - \frac{e}{m}\tau_0(F^{\mu\nu}a_\nu-(uFa)u^\mu) = \frac{e}{m}F^{\mu\nu}u_\nu \, .
\end{equation}
The quantity $uFa$ can be evaluated as
\begin{equation}
 uFa = u_\mu F^{\mu\nu} a_\nu = \frac{\mathcal{B}}{c^2}(\dot{u}_xu_y-u_x\dot{u}_y) \, .
\end{equation}
We then write out the spatial components of~\req{eq:FO_B}
\begin{align}
 \left(\frac{1}{\Omega_B}+\tau_0\frac{u_x u_y}{c^2}\right)\dot{u}_x-\tau_0\left(1+\frac{u_x^2}{c^2}\right)\dot{u}_y &= u_y\,, \\
 \tau_0\left(1+\frac{u_y^2}{c^2}\right)\dot{u}_x + \left(\frac{1}{\Omega_B}-\tau_0\frac{u_xu_y}{c^2}\right)\dot{u}_y &= -u_x \, .
\end{align}
This is a linear system of equations in the acceleration components, which can be inverted to obtain
\begin{align}
 \label{eq:uxdotBfield}\dot{u}_x &= \Omega_B\frac{u_y-\Omega_B\tau_0u_x\gamma^2}{1+(\Omega_B\tau_0\gamma)^2}\,, \\
 \label{eq:uydotBfield}\dot{u}_y &= -\Omega_B\frac{u_x+\Omega_B\tau_0u_y\gamma^2}{1+(\Omega_B\tau_0\gamma)^2}\,,
\end{align}
in which we used the constraint $u^2=c^2$ to simplify the equations.

We can also differentiate this constraint to relate $\dot{\gamma}$ to the above acceleration components~\req{eq:uxdotBfield} ,~\req{eq:uydotBfield}, as 
\begin{equation} \label{eq:gammadot}
 \dot{\gamma} = \frac{1}{\gamma}(u_x\dot{u}_x+u_y\dot{u}_y) \, .
\end{equation}
If we insert~\req{eq:uxdotBfield} and~\req{eq:uydotBfield} into~\req{eq:gammadot}, we get 
\begin{equation} \label{eq:gammadotBfield}
 \dot{\gamma} = \frac{\Omega_B^2\tau_0\gamma(1-\gamma^2)}{1+(\Omega_B\tau_0\gamma)^2} \, .
\end{equation}
We now have each component of the acceleration and, after a bit of manipulation, the square of its magnitude reads
\begin{equation}
 a_\mu a^\mu = -\frac{\Omega_B^2c^2(\gamma^2-1)}{1+(\Omega_B\tau_0\gamma)^2} \, ,
\end{equation}
which agrees exactly with~\req{eq:a2resummed}. We can then investigate the strong field limit $\Omega_B\rightarrow \infty$ to see
\begin{equation}
a_\mu a^\mu \rightarrow -\frac{c^2(\gamma^2-1)}{\tau_0^2\gamma^2} = -\frac{c^2}{\tau_0^2}\beta^2 \, .
\end{equation}
For ultrarelativistic particles, $\beta \rightarrow 1$, and we obtain a finite limit to the acceleration magnitude 
\begin{equation}
 a_\mu a^\mu \rightarrow -\frac{c^2}{\tau_0^2} \, .
\end{equation}
We see in this key result that the covariant limit to acceleration appears for a (magnetic) force acting, in this case, normal to the direction of motion.

We can also compare the EFO equation with the LL equation for the constant magnetic field. Written out in components, the LL acceleration is 
\begin{align}
 \dot{u}_x &= \Omega_B(u_y-\Omega_B\tau_0u_x\gamma^2)\,, \\
 \dot{u}_y &= -\Omega_B(u_x+\Omega_B\tau_0u_y\gamma^2)\,, \\
 \dot{\gamma} &= -\Omega_B^2\tau_0\gamma(\gamma^2-1) \, .
\end{align}
A study of the numerical properties of the LL system in a constant magnetic field is presented in~\cite{Elkina:2014dza}, and analytical solutions of the LL equation in constant fields appear in~\cite{Parga:1999, Yaremko:2013}. We note that each component of the LL acceleration is equivalent to the numerator in the corresponding EFO accelerations,~\req{eq:uxdotBfield},~\req{eq:uydotBfield}, and~\req{eq:gammadotBfield}. As the LL equation is a perturbative approximation to the EFO equation, we can see that the LL approximation is only valid for $\Omega_B\tau_0\gamma << 1$. Additionally, it is clear that the LL acceleration is unbounded as each component of the acceleration scales quadratically with the magnetic field. 

\begin{figure}
	\includegraphics[width=\linewidth]{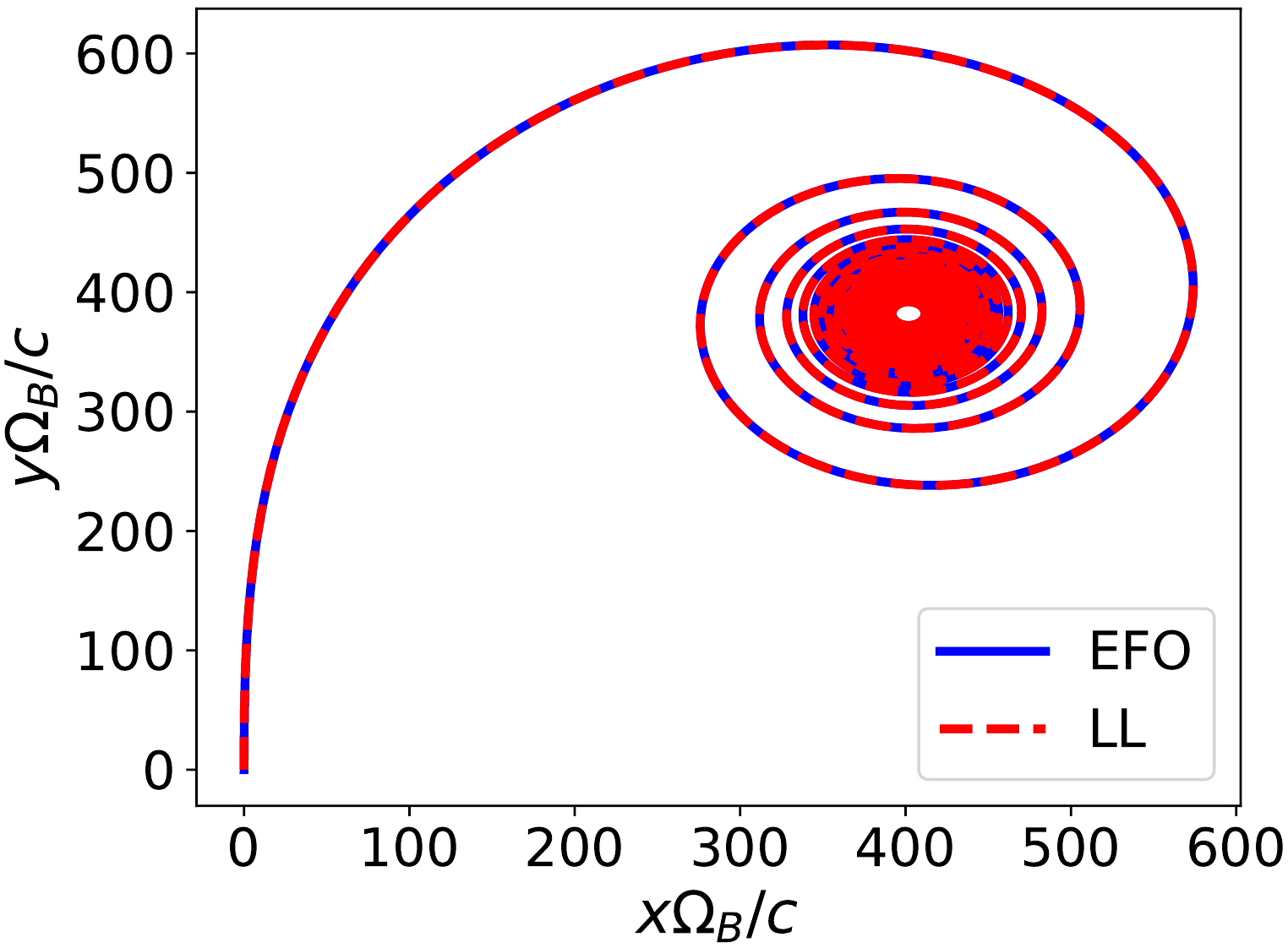}
	\includegraphics[width=\linewidth]{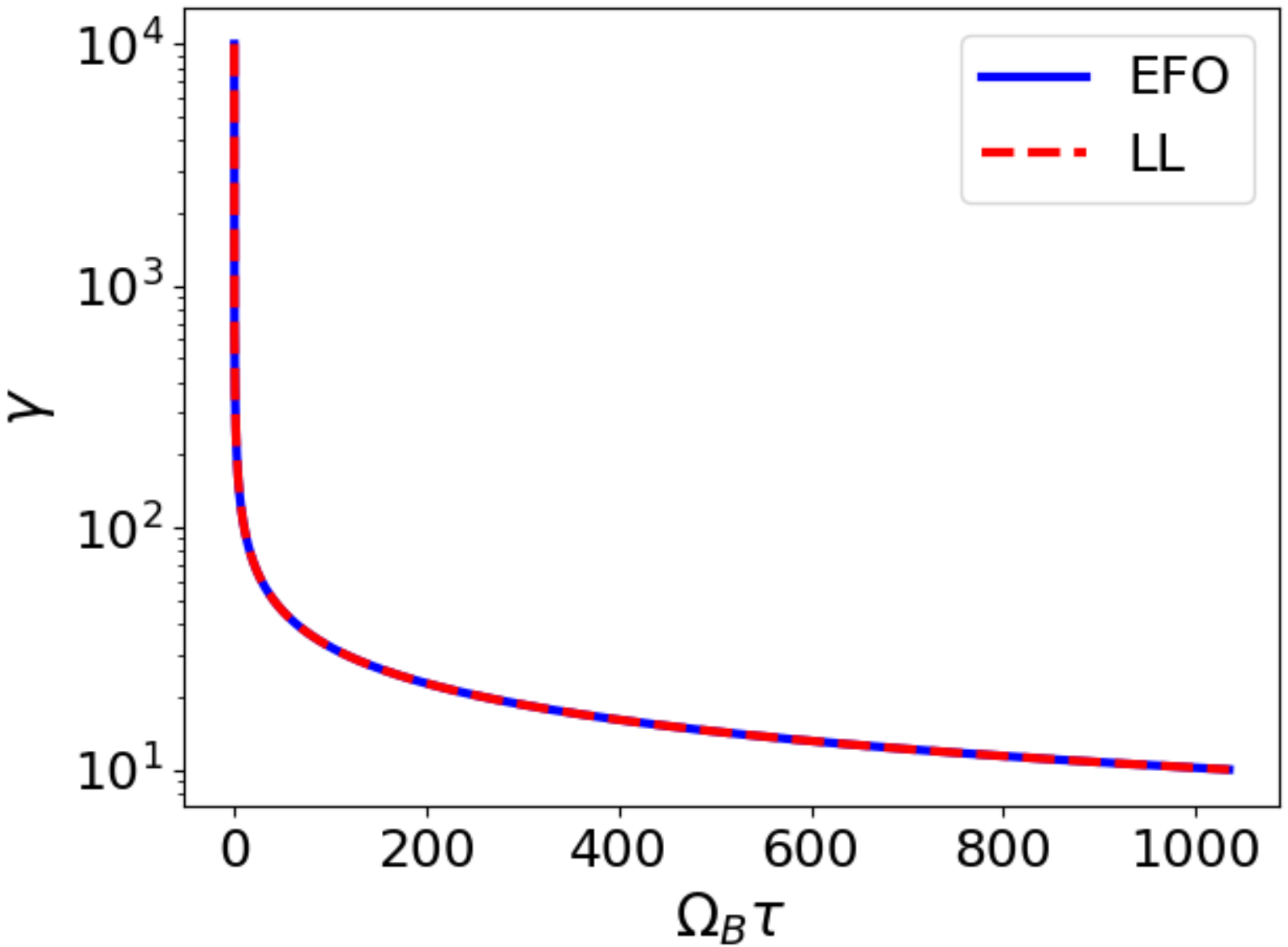}
	\caption{\label{fig:Bfield1} Numerical solutions of the EFO and LL equations for an electron in a subcritical field $\mathcal{B}= 4.41 \times 10^6 \, T = 10^{-3}B_c$ with an initial velocity corresponding to $\gamma_0=10^4$. The EFO curve is the solid blue line while the LL curve is the dashed red line. The plot above shows the spiraling trajectory of the particle, while the plot below shows $\gamma$ as a function of $\tau$. At this field strength the EFO and LL solutions are effectively identical.}
\end{figure} 

Numerical solutions for both the EFO and LL equations of motion are presented for an electron injected into the field with $\gamma_0=10^4$ at $\tau=0$. Figure \ref{fig:Bfield1} shows the motion of the particle in a weak (subcritical) field of $\mathcal{B}=4.41 \times 10^6 \, T = 10^{-3}B_c$ where $B_c$ is the Schwinger critical magnetic field discussed in Sec. \ref{Sec:Final}. At this field strength the EFO and LL equations yield nearly identical results.

\begin{figure}
 \includegraphics[width=\linewidth]{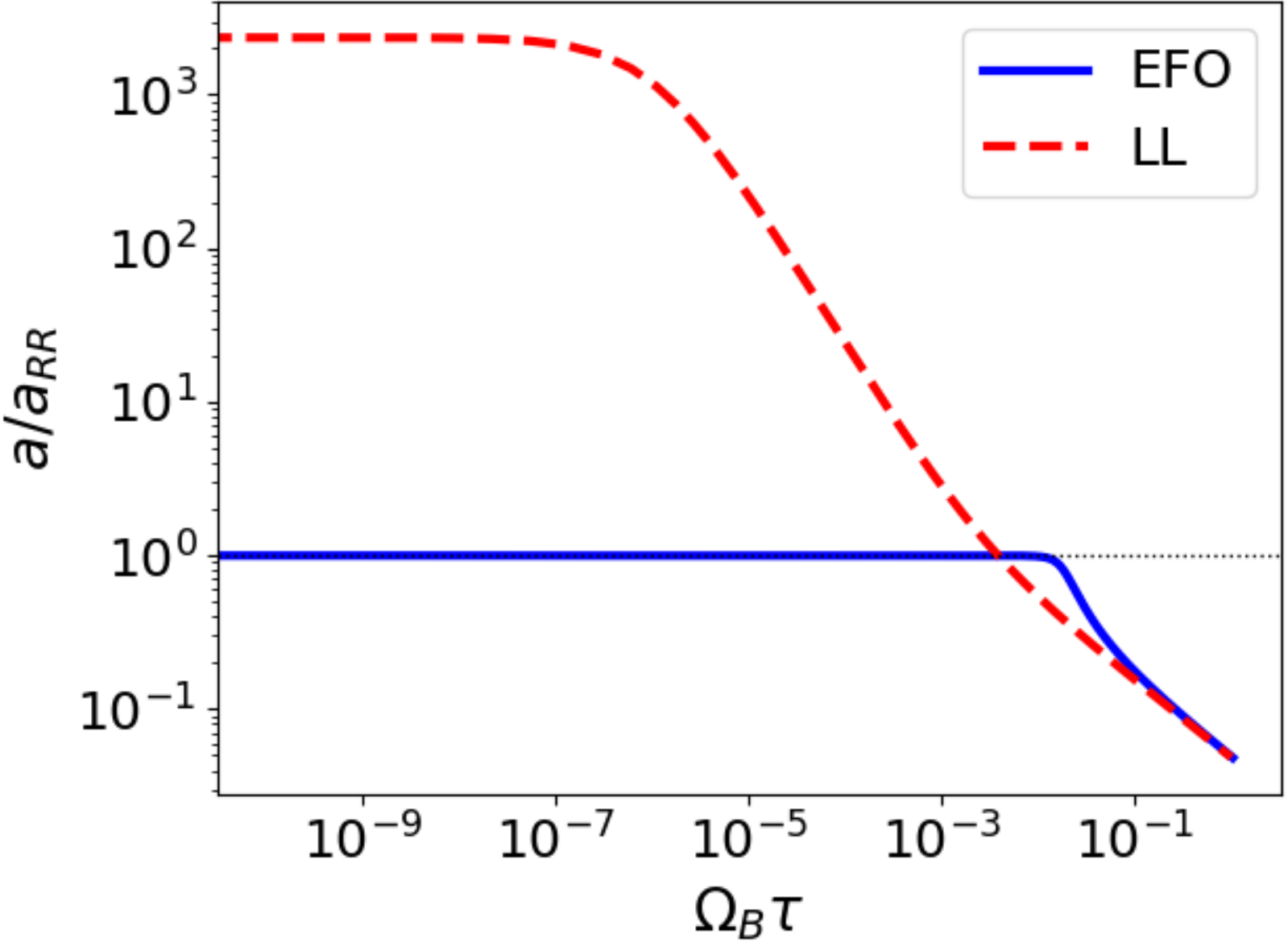}
	\includegraphics[width=\linewidth]{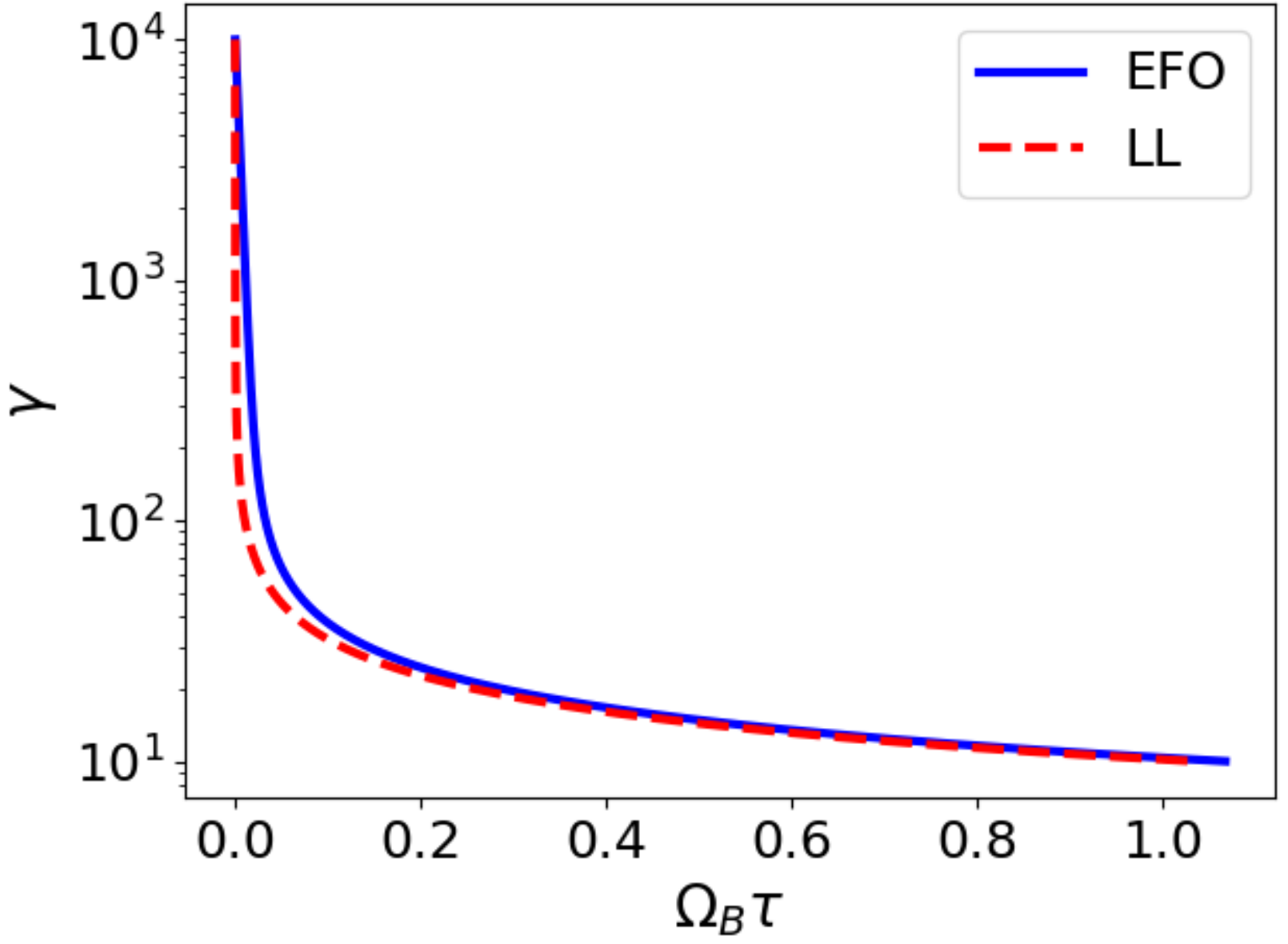}
	\caption{\label{fig:Bfield2} Numerical solutions of the EFO and LL equations for an electron in a critical field $\mathcal{B}= 4.41 \times 10^9 \, T = B_c$ with an initial velocity corresponding to $\gamma_0=10^4$. The plot above is of the invariant acceleration magnitude of the electron in units of $a_{\text{RR}}$. The plot below shows $\gamma$ as a function of $\tau$. The EFO and LL solutions differ above the limiting acceleration region, but quickly converge as the acceleration decreases.}
\end{figure}

Figure \ref{fig:Bfield2} shows the invariant magnitude of acceleration $a=\sqrt{-a_\mu a^\mu}$ and $\gamma$ for the EFO and LL equations. For the EFO solution, the particle begins with a value of $\gamma$ large enough to reach limiting acceleration, which is marked by the horizontal line. It radiates at a constant rate corresponding to this limiting acceleration until $\gamma$ decreases enough that the acceleration falls below $c/\tau_0$.The LL acceleration is not limited. The LL and EFO solutions differ only for short times while the acceleration is above or at the limiting value, and quickly converge when the acceleration drops below this limit. We note that at this field strength the LL equation is not valid since the perturbative approach is no longer justified. 

\subsection{Constant electric field ($\mathcal{S}<0, \, \mathcal{P}=0$)}\label{sec:Econstarbit}
We now consider a constant electric field in the $z$-direction so that $\pmb{\mathcal{E}}=(0,0,\mathcal{E})$. For motion parallel to the field, the Eliezer-Ford-O'Connell radiation reaction force vanishes as seen in section \ref{sec:EConst}. However, we can study a particle with initial velocity perpendicular to the field so that the motion is in two dimensions. The particle will turn into the field and radiate while doing so. In this case the particle will have a 4-velocity 
\begin{equation}
 u^\mu = (\gamma c , u_\parallel, 0, u_\perp) = \gamma c(1, \beta_\parallel, 0, \beta_\perp)\,,
\end{equation}
where $u_\parallel$, $u_\perp$ are the components of the particle's velocity parallel and perpendicular to the field, respectively. The components of the acceleration for the EFO equation can be solved for to obtain
\begin{align}
 \dot{u}_\parallel &= \Omega_E \frac{\gamma c-\Omega_E\tau_0 u^2_\perp u_\parallel/c^2}{1+(\Omega_E\tau_0 u_\perp/c)^2}\,, \\
 \dot{u}_\perp &= -\Omega_E^2\tau_0 \frac{u_\perp(1+u^2_\perp/c^2)}{1+(\Omega_E \tau_0 u_\perp/c)^2}\,, \\
 \dot{\gamma} &= \Omega_E\frac{u_\parallel/c-\Omega_E\tau_0 u^2_\perp\gamma/c^2}{1+(\Omega_E\tau_0u_\perp/c)^2} \, .
\end{align}
Using these components to calculate the square of the acceleration results in 
\begin{equation} \label{eq:asquared_Efield}
 a_\mu a^\mu = -\frac{\Omega_E^2 c^2(1+u_\perp/c)^2}{1+(\Omega_E\tau_0u_\perp/c)^2} \, .
\end{equation}
Taking the strong field limit $\Omega_E \rightarrow \infty$ of~\req{eq:asquared_Efield}, we arrive at 
\begin{equation}
 a_\mu a^\mu \rightarrow -\left(\frac{c}{\tau_0}\right)^2 \frac{(1+u_\perp/c)^2}{u_\perp^2/c^2} \, .
\end{equation}
We then see that the existence of limiting acceleration in this case is dependent on the transverse velocity $u_\perp$. We can separately take the limits of large and small transverse velocity to obtain
\begin{equation}
 a_\mu a^\mu \rightarrow \begin{cases} -c^2/\tau_0^2, & \beta_\perp \rightarrow 1, \\
 -\infty, & \beta_\perp \rightarrow 0. \end{cases}
\end{equation}
While the particle maintains an ultrarelativistic speed in the direction transverse to the field, its acceleration will be bounded. However, for a particle moving parallel to the field, the acceleration is unbounded as we know the EFO radiation reaction force vanishes.

Again, we can compare the Eliezer-Ford-O'Connell solution with that of the Landau-Lifshitz equation. Writing out each component of the acceleration, we have 
\begin{align}
 \dot{u}_\parallel &= \Omega_E(\gamma c - \Omega_E\tau_0 u^2_\perp u_\parallel/c)\,, \\
 \dot{u}_\perp &= - \Omega_E^2\tau_0 u_\perp (1+u^2_\perp/c^2)\,, \\
 \dot{\gamma} &= \Omega_E(u_\parallel/c - \Omega_E\tau_0 u^2_\perp \gamma/c^2) \,.
\end{align}
We can see that the LL acceleration components are equivalent to the EFO acceleration components expanded to order $\tau_0$.
\subsection{Constant crossed fields ($\mathcal{P}=0$)}
We now consider constant crossed electric and magnetic fields, $\pmb{\mathcal{E}}=(\mathcal{E},0,0)$ and $\pmb{\mathcal{B}}=(0,0,\mathcal{B})$. We then can study the following regimes: electrically dominated ($\mathcal{S}<0$), magnetically dominated ($\mathcal{S}>0$), and low-frequency plane wave ($\mathcal{S}=0$) regime. We note that the cases of $\mathcal{S}<0$ and $\mathcal{S}>0$ are related to the pure constant electric (Sec.~\ref{sec:Econstarbit}) and magnetic fields (Sec.~\ref{sec:Bconst}) respectively by a Lorentz transformation. The case of $\mathcal{S}=0$ is the only new configuration arising in discussion of the long-wavelength limit of plane waves. Since constant crossed fields are important for the locally constant field approximation we will consider all possible values of $\mathcal{S}$ in this section. 

Let the particle motion be restricted to the $xy$ plane so that its 4-velocity is
\begin{equation}
 u^\mu = (\gamma c, u_x, u_y, 0) = \gamma c(1, \beta_x, \beta_y, 0) \, . 
\end{equation} 
We will now evaluate our expression~\req{eq:a2resummed} to determine under which circumstances a limiting acceleration occurs. The Lorentz force acceleration \req{eq:a2lf} can be written as
\begin{equation}
 a^2_{LF} = c^2 (\Omega_B^2 -\Omega_E^2 - (\Omega_E u_y/c + \Omega_B \gamma)^2) \, .
\end{equation}
From this, we can then evaluate the EFO acceleration using~\req{eq:a2resummed}
\begin{equation}
 a_\mu a^\mu = \frac{c^2 (\Omega_B^2 -\Omega_E^2 - (\Omega_E u_y/c + \Omega_B \gamma)^2)}{1+\tau_0^2(\Omega_E u_y/c+\Omega_B \gamma)^2} \, .
\end{equation}

We will first consider $\mathcal{S}\geq 0$, which encompasses both the magnetically dominated and low-frequency plane wave cases. We can rewrite our acceleration as 
\begin{equation}
 a_\mu a^\mu = c^2 \frac{(\Omega_B^2-\Omega_E^2-(\Omega_E u_y/c + \Omega_B \gamma)^2)}{1+\tau_0^2(\Omega_E u_y/c+\Omega_B \gamma)^2} \, .
\end{equation}
Since we must have $a^2<0$, as the acceleration is a spacelike vector, we can conclude that the second term in the numerator must always be larger than the first. This term will therefore dominate in the limit $\Omega_E, \Omega_B \rightarrow \infty$, and we see that 
\begin{equation}
 a_\mu a^\mu \rightarrow -c^2 \frac{(\Omega_E u_y/c + \Omega_B \gamma)^2}{\tau_0^2(\Omega_E u_y/c+\Omega_B \gamma)^2} \rightarrow -\left(\frac{c}{\tau_0}\right)^2 \, .
\end{equation}
Therefore the acceleration will be bounded for $\mathcal{S}\geq 0$. 
We now turn to the electrically dominated case, $\mathcal{S}<0$. If we first consider when $u_y=0$ so that the particle is moving parallel to the electric field, we have
\begin{equation}
 a_\mu a^\mu = -c^2 \frac{\Omega_E^2+\Omega_B^2(\gamma^2-1)}{1+(\tau_0\Omega_B\gamma)^2} \, .
\end{equation}
If we take the limit in which $\Omega_E$, $\Omega_B \rightarrow \infty$, we see that the acceleration will reach the limiting value $c/\tau_0$ only if $\Omega_B \gamma >> \Omega_E$. That is, if the magnetic field is strong enough to deflect the particle away from motion parallel to the electric field, then the acceleration will be bounded. Otherwise, the particle will continue to travel parallel to the electric field with an unbounded acceleration. 

If we now consider when the particle's velocity is entirely in the transverse direction, $u_y \approx \gamma c$, we have in the strong field limit
\begin{equation}
 a_\mu a^\mu \rightarrow -c^2 \frac{\Omega_E^2 + \Omega_E^2 \gamma^2}{1+\tau_0^2\Omega_E^2\gamma^2} \rightarrow - \left(\frac{c}{\tau_0}\right)^2 \, .
\end{equation}
The acceleration is therefore bounded for a particle moving transverse to the electric field, and it will only become unbounded if it is allowed to turn entirely in the direction parallel to the electric field. 

From the discussion in this section we see that all cases of an acceleration greater than $c/\tau_0$ are rooted in the problem of a particle moving parallel to a constant electric field. For $\mathcal{S}\geq 0$ the magnetic field is strong enough to keep this from happening, but the particle can turn entirely in the direction of the electric field for $\mathcal{S}<0$.

\subsection{Constant parallel fields of the same magnitude ($\mathcal{S}=0, \, \mathcal{P} \neq 0)$}
We now consider the case of constant parallel electric and magnetic fields of equal strength, both pointing in the $z$-direction: $\pmb{\mathcal{E}}=(0,0,\mathcal{E})$ and $\pmb{\mathcal{B}}=(0,0,\mathcal{B})$. For fields of equal magnitude, we have
\begin{equation}
 \Omega_E = \Omega_B \equiv \Omega \, .
\end{equation}
Let the particle's 4-velocity have a component along each axis
\begin{equation}
 u^\mu = (\gamma c, u_x, u_y, u_z) \, .
\end{equation}
We will again use~\req{eq:a2resummed} to determine if acceleration is limited for this case. First evaluating the Lorentz force acceleration, we find
\begin{equation}
 a_{LF}^2 = -c^2 \Omega^2 [(\gamma^2 - u_z^2/c^2)+(u_x^2 + u_y^2)/c^2] \, .
\end{equation}
Defining the transverse component of the velocity as 
\begin{equation}
 u_\perp^2 = u_x^2 + u_y^2 \,,
\end{equation}
the Lorentz force acceleration becomes 
\begin{equation}
 a_{LF}^2 = -c^2\Omega^2(1+2u_\perp^2/c^2) \, .
\end{equation}
This allows us to write the EFO acceleration as 
\begin{equation}
 a_\mu a^\mu = -\frac{c^2\Omega^2(1+2u^2_\perp/c^2)+c^2\tau_0^2\Omega^4}{1+\tau_0^2\Omega^2(1+2u^2_\perp/c^2)} \, .
\end{equation}
We now take the strong field limit $\Omega\rightarrow \infty$. If we first consider $u_\perp=0$, we again have the case of a particle moving parallel to a constant electric field. We already know that the acceleration reduces to its Lorentz force value,
\begin{equation}
 a_\mu a^\mu = -c^2 \Omega^2 \rightarrow -\infty \, ,
\end{equation}
which is unbounded. This result is obtained by canceling the terms in the numerator and denominator. For $u_\perp \neq 0$, the presence of the $\mathcal{P}$ term in this example again leads to an unbounded acceleration:
\begin{equation}
 a_\mu a^\mu \rightarrow -\frac{c^2\Omega^2}{1+2u_\perp^2/c^2} \rightarrow -\infty \, .
\end{equation}
We can infer that field configurations with a nonzero value of $\mathcal{P}$ do not yield a limiting acceleration.

\section{Light Plane Wave Field}\label{sec:lightM}
The plane wave field is our first example of a time-dependent field where the $u\cdot \partial F^{\mu\nu}u_\nu$ term in both the LL and EFO equations will be relevant since $\mathcal{S}=0$ and $\mathcal{P} = 0$. We consider a plane wave field 4-potential 
\begin{equation}\label{eq:planewave}
A^\mu = \varepsilon^\mu \mathcal{A}_0 f(\xi), \qquad \xi = k \cdot x\,,
\end{equation}
where $f$ is an arbitrary function of the phase $\xi$ and $k^\mu$ is the lightlike wave vector with $k^2 = 0$ orthogonal to polarization $\varepsilon^\mu$ so that $k \cdot \varepsilon = 0$ and $\varepsilon^2 = -1$. Then for the EM tensor~\req{eq:EMtensor} and its derivative we have
\begin{align}
F^{\mu\nu} &= (k^\mu\varepsilon^\nu - \varepsilon^\mu k^\nu)f'(\xi)\mathcal{A}_0,\\
\dot{F}^{\mu\nu} &= (k^\mu\varepsilon^\nu - \varepsilon^\mu k^\nu)(k \cdot u)f''(\xi)\mathcal{A}_0\,.
\end{align}
The primes denote derivatives with respect to the phase $\xi$. Substituting into the Eliezer-Ford-O'Connell equation of motion~\req{eq:FO2}, we get
\begin{multline}\label{eq:plane_wave_motion}
a^\mu = \frac{e}{m}[k^\mu (\varepsilon \cdot u) - \varepsilon^\mu (k \cdot u)][f' + \tau_0 (k \cdot u)f'']\mathcal{A}_0 \\
+ \tau_0 \frac{e}{m}[k^\mu (\varepsilon \cdot a) - \varepsilon^\mu (k \cdot a)]f' \mathcal{A}_0\\
- \tau_0 \frac{e}{m} \frac{u^\mu}{c^2}[(k \cdot u)(\varepsilon \cdot a) - (\varepsilon \cdot u)(k \cdot a)]f' \mathcal{A}_0\,.
\end{multline}
If we multiply by $a_\mu$ we obtain
\begin{equation}\label{eq:a2hybrid}
a^2 = \frac{e\mathcal{A}_0}{m} [f' + \tau_0 (k\cdot u)f'']W\,,
\end{equation}
where we defined the tensor part of the contraction $uFa$ as
\begin{equation}
W \equiv (k\cdot a)(\varepsilon \cdot u) - (\varepsilon \cdot a)(k \cdot u)\,.
\end{equation}
If we project the equation of motion~\req{eq:plane_wave_motion} by $k_\mu$ or $\varepsilon_\mu$ we get, respectively,
\begin{equation}
\label{eq:ka}k \cdot a = \tau_0 \frac{e\mathcal{A}_0}{m}\frac{k \cdot u}{c^2}f'W\,,
\end{equation}
\begin{multline}
\label{eq:ea}\varepsilon \cdot a = \frac{e\mathcal{A}_0}{m}(k \cdot u)[f' + \tau_0 (k\cdot u) f''] \\
+\tau_0 \frac{e\mathcal{A}_0}{m}(k\cdot a) f' + \tau_0 \frac{e\mathcal{A}_0}{m} \frac{\varepsilon \cdot u}{c^2}f' W\,.
\end{multline}
Now we eliminate $W$ by evaluating the combination~\req{eq:ka}$\times(\varepsilon \cdot u)$ -~\req{eq:ea}$\times(k \cdot u)$ and substitute for $k \cdot a$ from~\req{eq:ka}. Thus $W$ is equal to
\begin{equation}\label{eq:W}
W = \frac{- \frac{e\mathcal{A}_0}{m}(k\cdot u)^2 [f' + \tau_0 f'' (k \cdot u)]}{1 + \tau_0^2 \frac{e^2\mathcal{A}_0^2}{m^2c^2}(k \cdot u)^2 f'^2}\,,
\end{equation}
which we can substitute back to~\req{eq:a2hybrid}
\begin{equation} \label{eq:a2_wave}
a_\mu a^\mu = \frac{- \frac{e^2\mathcal{A}_0^2}{m^2}(k\cdot u)^2 [f' + \tau_0 f'' (k \cdot u)]^2}{1 + \tau_0^2 \frac{e^2\mathcal{A}_0^2}{m^2c^2}(k \cdot u)^2 f'^2}\,.
\end{equation}
We can write~\req{eq:a2_wave} in terms of the Lorentz force acceleration, which is
\begin{equation}
 a^2_\text{LF} = -\frac{e^2\mathcal{A}_0^2}{m^2}(k\cdot u)^2f'^2 \, ,
\end{equation}
giving us a final result of 
\begin{equation}\label{eq:a2EFOplanewave}
 a_\mu a^\mu = \frac{a_{LF}^2[1+\tau_0(k\cdot u)f''/f']^2}{1+\frac{\tau_0^2}{c^2}|a_{LF}^2|} \, .
\end{equation}
Compared to the first iteration of the Landau-Lifshitz model, which has an analytical solution for plane wave fields~\cite{Dipiazza2008,Hadad:2010mt,Yaghjian:2021}, the expression for $k\cdot a$~\req{eq:ka} and \req{eq:W} in our model has a complicated dependence on $k\cdot u$, and thus, is in general, not integrable. 

We notice that if the $f'$ term in the numerator of~\req{eq:a2_wave} dominates over the $f''$ term in the limit of strong fields, the acceleration will approach the usual limiting value $a^2 \rightarrow -(c/\tau_0)^2$. However, the $f''$ term will be large if 
\begin{equation}
 \tau_0\omega\gamma \sim 1 \, .
\end{equation}
If we take as an example a wavelength of $1000$ nm, then $\omega\tau_0 \sim 10^{-8}$ for an electron. Overcoming the limiting acceleration in this case would take an electron of energy 50 TeV. As shown in~\cite{Kravets:2013}, a particle colliding with a wave will typically radiate energy too quickly to reach large accelerations. We can conclude that an electron in a plane wave of typical experimental amplitude and frequency will not reach above the value of the limiting acceleration.

For comparison, we can compute the LL acceleration squared \req{eq:a2_LL} as
\begin{equation} \label{eq:a2_wave_LL}
 a_\mu a^\mu = a_{LF}^2\left\{[1+\tau_0(k\cdot u)f''/f']^2+\frac{\tau_0^2}{c^2}|a_{LF}^2|\right\} \, .
\end{equation}
If we expand the result~\req{eq:a2EFOplanewave} in the powers of $\tau_0$, we can match the LL result \req{eq:a2_wave_LL} in linear order and the expressions again start to differ for quadratic order and higher as we saw in constant fields. 

\section{Application for the material friction force}\label{sec:material_friction}
As we have shown in~\cite{Formanek:2020zwc}, studying a material friction as the external force driving the motion poses certain conceptual advantages. The material medium provides a unique reference frame which is not available in the Lorentz-invariant vacuum. Also, the use of an empirical friction force avoids the need to resolve the consistency issue between the EM field equations and a description of charged particle motion. Finally, since the motion in the rest frame of the medium is in one dimension, it poses a simple case for which the RR force does not vanish and the EFO equation can be inverted to obtain $a^\mu$ rather than just its magnitude.

The external material friction force can be taken in covariant form as
\begin{equation}
\mathcal{F}^\mu_\text{ext} = mr R^{\mu\nu}u_\nu\,,
\end{equation}
where $r$ is the resistive medium friction coefficient and the antisymmetric tensor $R^{\mu\nu}$ reads
\begin{equation}\label{eq:Rmunu}
R^{\mu\nu} = \eta^\mu u^\nu - u^\mu \eta^\nu\,,
\end{equation} 
where $\eta^\mu$ is the constant 4-velocity of the resistive medium with $\eta^2 = c^2$. The Eliezer-Ford-O'Connell formulation~\req{eq:FO} for this driving force leads to
\begin{equation}
a^\mu_\mathrm{EFO} = r R^{\mu\nu}u_\nu + \tau_0 r P^\mu_\alpha \frac{d}{d\tau}(R^{\alpha \nu}u_\nu)\,.
\end{equation}
We use the subscript EFO to avoid confusion with the LL and Mo-Papas (MP) models, which are also discussed in this section. If we substitute the tensor $R^{\mu\nu}$~\req{eq:Rmunu} we obtain 
\begin{equation}
a^\mu_\mathrm{EFO} = r R^{\mu\nu}u_\nu - \tau_0 r (\eta \cdot u)a^\mu_\mathrm{EFO}\,.
\end{equation}
We can then solve for the 4-acceleration to get
\begin{equation}\label{eq:amu_ours}
a^\mu_\mathrm{EFO} = \frac{rR^{\mu\nu}u_\nu}{1+\tau_0 r (\eta \cdot u)}\,.
\end{equation}
Defining the acceleration due to solely the material friction force as
\begin{equation}
a^\mu_\text{ext} = rR^{\mu\nu}u_\nu \, ,
\end{equation}
we can write the invariant square of the full acceleration as
\begin{equation}\label{eq:a2medfric}
a^2_\mathrm{EFO} = \frac{a_\text{ext}^2}{[1+\tau_0 r (\eta \cdot u)]^2}\,.
\end{equation}
The effect of the radiation reaction is to lower the particle's acceleration. The $\eta \cdot u$ product is manifestly positive
\begin{equation}
\eta \cdot u = \gamma_\text{med} \gamma c^2(1-
\pmb{\beta}_\text{med} \cdot \pmb{\beta})\,,
\end{equation}
where the ``med'' subscript refers to quantities belonging to the medium. In the rest frame of the medium $\gamma_\text{med} = 1$ and $\pmb{\beta}_\text{med} = 0$, so $\eta \cdot u = c\gamma $. In the rest frame of the medium the motion is also in one dimension (see~\cite{Formanek:2020zwc}) and if we take the $r\rightarrow \infty$ and $\beta\rightarrow 1$ limits of~\req{eq:a2medfric} we obtain
\begin{equation}
a^2_\mathrm{EFO} \rightarrow - \frac{c^2}{\tau_0^2}\,,
\end{equation}
which is the same limiting acceleration as in the case of an external electromagnetic force. We can also define an effective mass 
\begin{equation}
 M(\eta\cdot u,\,r) \equiv 1+\tau_0 r(\eta\cdot u)\,,
\end{equation}
which is dependent both on the relative velocity between the medium and particle as well as the strength of the friction force. The equation of motion then takes the Newtonian-like form
\begin{equation}
 M(\eta\cdot u,\,r)a^\mu_\mathrm{EFO} = mrR^{\mu\nu}u_\nu \, .
\end{equation}
We can then interpret the effect of the Eliezer-Ford-O'Connell radiation reaction force as a change in the inertia of the particle. When the limiting acceleration $c/\tau_0$ is reached, the force goes entirely into increasing the effective mass rather than accelerating the particle. 
We can contrast our result with the Mo-Papas formulation which omits the proper time derivative of the $R^{\mu\nu}$ term
\begin{equation}
a^\mu_\mathrm{MP} = rR^{\mu\nu}u_\nu + \tau_0 r P^\mu_\alpha R^{\alpha}_\nu a^\nu_\mathrm{MP}\,,
\end{equation}
where the second term identically vanishes since
\begin{equation}
R^{\alpha}_\nu a^\nu_\mathrm{MP} = - u^\mu (\eta \cdot a_\mathrm{MP})
\end{equation}
is proportional to 4-velocity which gets projected out by $P^\mu_\alpha$. Thus there is no radiation friction contribution $a^\mu_\mathrm{MP} = a^\mu_\text{ext}$. For the Eliezer-Ford-O'Connell equation the entire radiation reaction effect comes from the $\frac{d}{d\tau} R^{\mu\nu}$ term.

We then turn to the Landau-Lifshitz equation with the material friction force. The first order iteration of the Landau-Lifshitz model has the form
\begin{equation}
a^\mu_\mathrm{LL} = rR^{\mu\nu}u_\nu + r\tau_0 \left[\frac{d}{d\tau}(R^{\mu\nu})u_\nu + rP^{\mu}_\alpha R^{\alpha\beta}R_{\beta\gamma}u^\gamma \right]\,.
\end{equation}
Again the last term vanishes, because 
\begin{equation}
R^{\alpha\beta}R_{\beta\gamma}u^\gamma = [(\eta\cdot u)^2 -c^4]u^\mu 
\end{equation}
is proportional to 4-velocity. If we apply the usual iteration scheme by substituting in the driving force for the acceleration, we get to lowest order
\begin{align}
a^\mu_\mathrm{LL} &= r R^{\mu\nu}u_\nu - \tau_0 r^2 (\eta \cdot u)R^{\mu\nu}u_\nu\\
\label{eq:a_LL_medium} &= [1- \tau_0 r (\eta \cdot u)]rR^{\mu\nu}u_\nu\,.
\end{align}
The square of the first iteration LL acceleration is 
\begin{equation}
a^2_\mathrm{LL} = a_\mathrm{ext}^2[1-\tau_0 r (\eta\cdot u)]^2\,.
\end{equation}
If we compare with the EFO result we again see agreement in linear order of $\tau_0$ and differences at second order and beyond.

In this simple 1D example we can continue the Landau-Lifshitz iterations to arrive at an alternating geometric series in $\tau_0 r (\eta\cdot u)$,
\begin{equation}
 a^\mu_{\text{LL},\infty} = a^\mu_\text{ext} \sum_{n=0}^\infty(-\tau_0r(\eta\cdot u))^n \, .
\end{equation}
If $\tau_0 r (\eta \cdot u) < 1$, we can resum the series and obtain the Eliezer-Ford-O'Connell result~\req{eq:amu_ours}. Thus, we have discovered the radius of convergence for the Landau-Lifshitz approximation for this problem. If the convergence condition is not satisfied, the Landau-Lifshitz equation will fail to yield physical solutions and will instead have only runaway solutions reminiscent of those of the LAD equation. This is made clear by noting that the sign of the acceleration changes in~\req{eq:a_LL_medium} when $\tau_0r(\eta\cdot u)>1$. This results in a particle being accelerated by the radiation reaction force rather than decelerated. It is notable that the EFO equation has no such deficiency for $\tau_0 r (\eta \cdot u) > 1$ and predicts that a particle will experience limiting acceleration in this regime. 

Interestingly, the resummation of the LL series yields the EFO equation of motion and not LAD. A more sophisticated resummation of the LL series is carried out in~\cite{Ekman:2021eqc} in the locally constant crossed field approximation. The resulting equation is found to differ from LAD for short times before the two converge. Since we directly obtain the EFO equation through resummation for the simple example of mechanical friction force, we can speculate that extending the results of~\cite{Ekman:2021eqc} to compare with the EFO equation may yield a closer relationship of the resummed LL equation with EFO than LAD.

For the purpose of solving the equations of motion numerically, we write them in terms of the rapidity 
\begin{equation}
 \cosh(y) = \gamma \, ,
\end{equation}
and a dimensionless friction strength 
\begin {equation}
 \tilde r = \frac{r\tau_0}{mc^2} \, . 
\end{equation}
Written in terms of these quantities and evaluated in the medium's rest frame, the EFO equation takes the form
\begin{equation}
 \frac{dy_\text{EFO}}{d\tau} = -\frac{\tilde{r} \sinh (y_\text{EFO})/\tau_0}{1+\tilde{r}\cosh(y_\text{EFO})} \, .
\end{equation}
Similarly, we can write the Landau-Lifshitz equation as 
\begin{equation} 
 \frac{dy_\text{LL}}{d\tau} = -\frac{\tilde{r}}{\tau_0} \sinh (y_\text{LL})[1-\tilde{r} \cosh (y_\text{LL})] \, .
\end{equation}
Here we can clearly see the range of validity of the LL equation. The force changes sign at $\tilde r\gamma = 1$; a positive force will accelerate, rather than decelerate the particle, resulting in behavior similar to the runaway solutions of the LAD equation~\cite{Rohrlich:1965-90}.

\begin{figure}
\includegraphics[width=\linewidth]{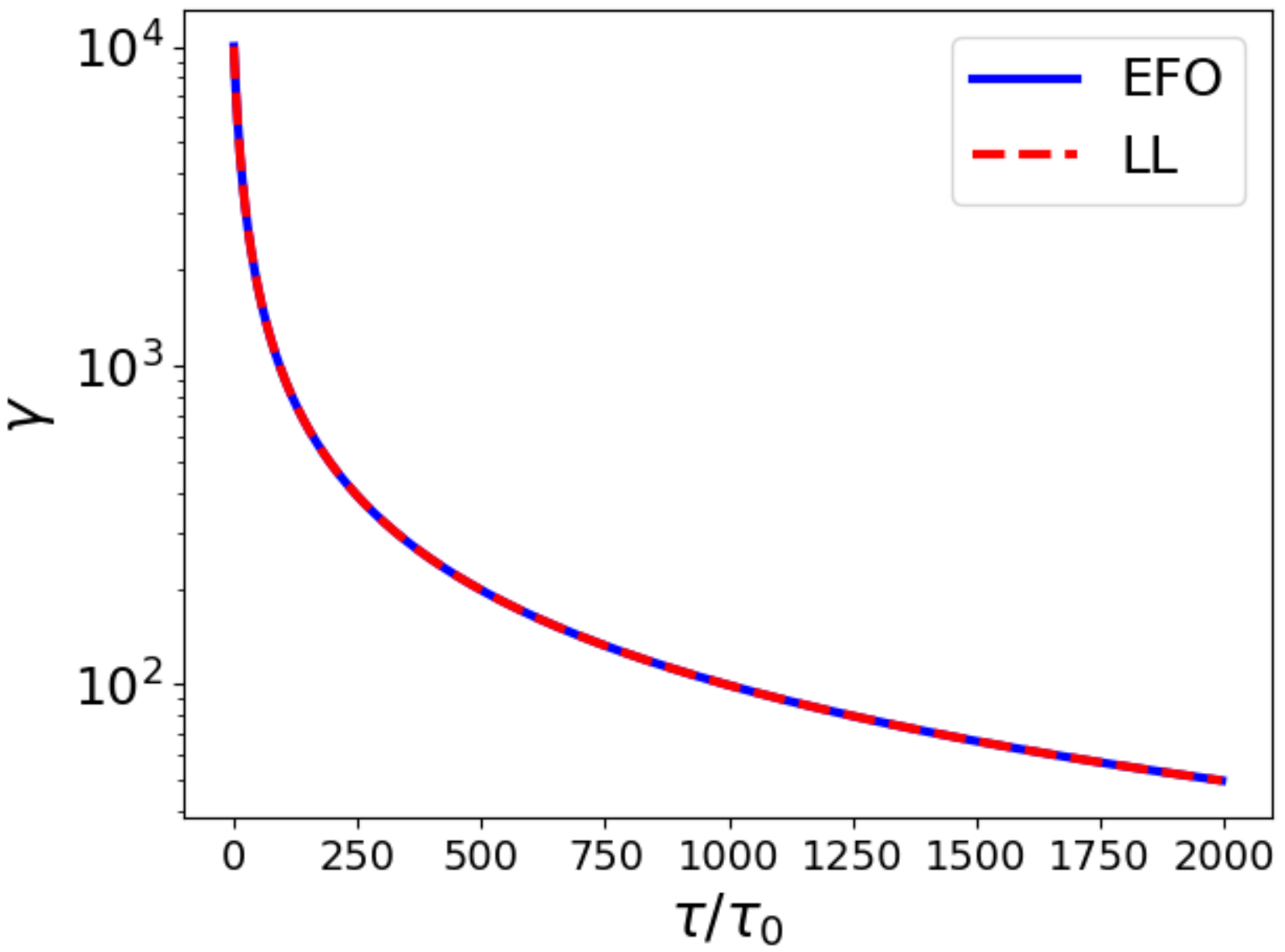}
\caption{\label{fig:material_friction1}$\gamma$ as a function of $\tau$ for the EFO and LL equations with $\tilde r = 10^{-5}$ and $\gamma_0=10^4$. $\tilde r \gamma < 1$ for the entirety of the motion so the LL and EFO equations are equivalent.}
\end{figure}

Figure \ref{fig:material_friction1} shows the weak acceleration behavior of $\gamma$ for the EFO and LL equations with $\tilde r=10^{-5}$. $\tilde{r}\gamma<1$, so the two equations give identical results. 

\begin{figure}
\includegraphics[width=\linewidth]{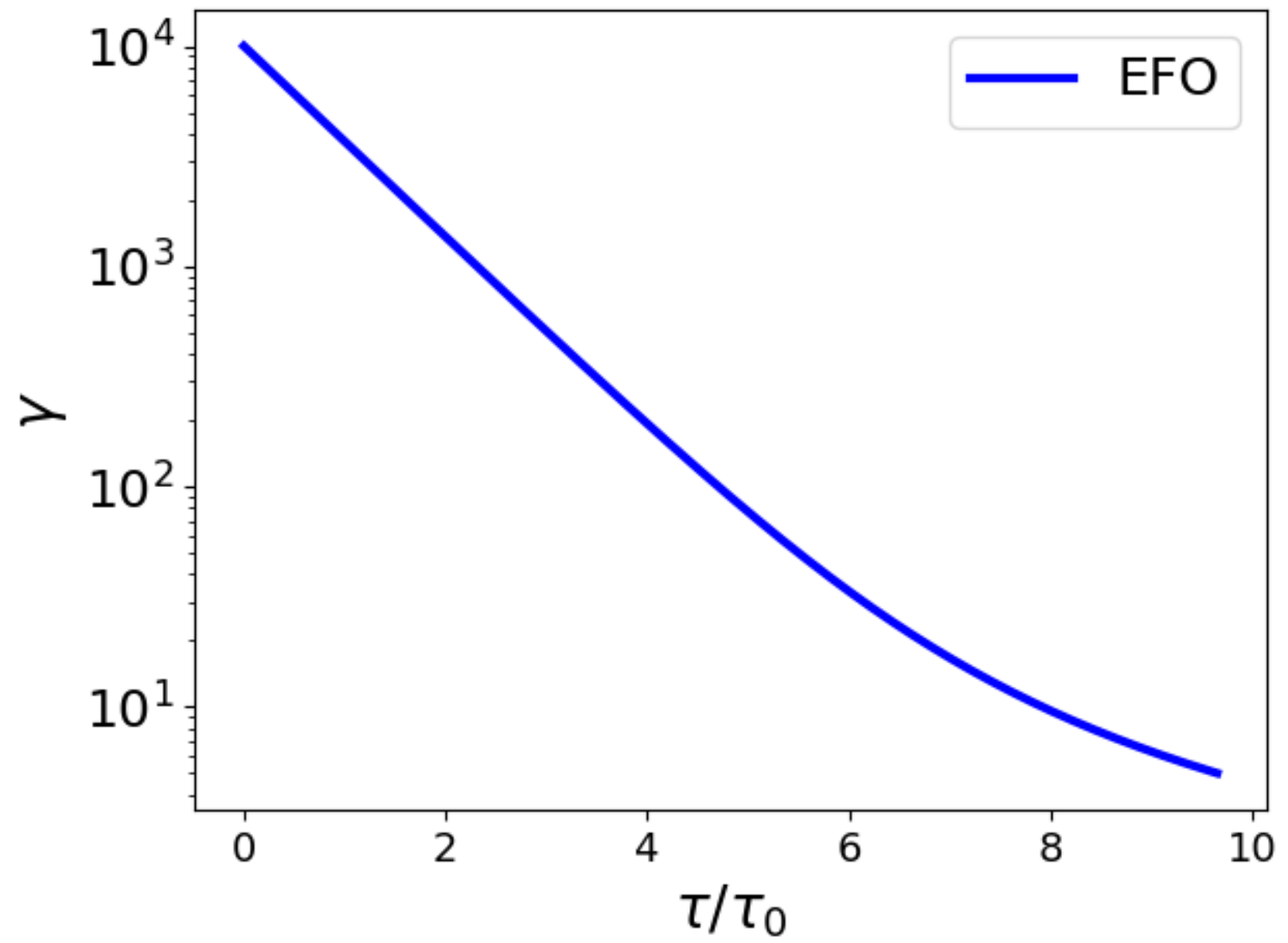}
\includegraphics[width=\linewidth]{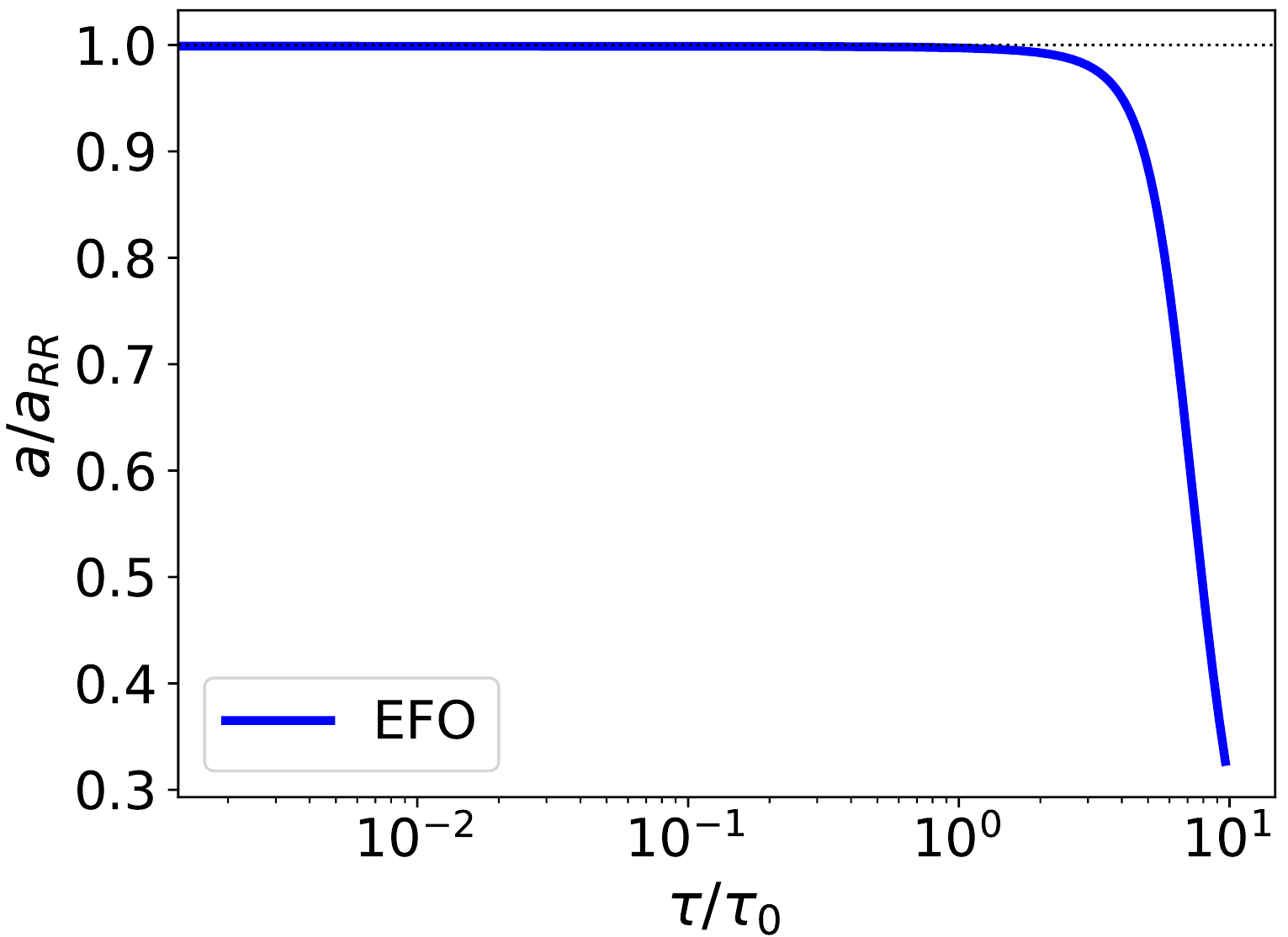}
\caption{\label{fig:material_friction2}EFO solution for $\tilde r = 10^{-3}$ and $\gamma_0=10^4$. The LL solution yields runaway solutions at this friction strength and is not shown on the plot. The plot above is of $\gamma$ as a function of $\tau$ while the plot below is of the invariant acceleration magnitude.}
\end{figure}

Figure \ref{fig:material_friction2} shows both $\gamma$ and the invariant acceleration magnitude for the EFO equation with $\tilde r = 10^{-3}$ and $\gamma_0=10^4$. Since $\tilde r \gamma > 1$, limiting acceleration is reached. The LL equation, on the other hand, breaks down for a friction force of this strength and predicts runaway solutions that rapidly accelerate instead of decelerate the particle.

\section{On a quantum limit to acceleration}\label{sec:quantum}
 
The extreme forces required to explore the limits to acceleration can be achieved today by, for example, colliding relativistic electrons with laser pulses~\cite{Hadad:2010mt}, provided that the classical electron dynamics allowing for radiation loss is introduced. This is commonly done in terms of the Landau-Lifshitz~\cite{LL:1962} field-dependent formulation of the Larmor radiation power formula, resulting in the equation of motion~\req{eq:LL}. When the LL perturbative approach is not applicable, the classical description of the interaction is usually superseded by the quantum framework. In this section we will discuss a limit on acceleration arising from the field screening by a pair production from vacuum.

Let us start by comparing the EFO limiting acceleration with the natural scales appearing in the quantum (QED) and classical (CED) descriptions of electrodynamics. In QED the commonly quoted ``critical'' Euler-Heisenberg-Schwinger field~\cite{Heisenberg:1936,Schwinger:1951nm} corresponds to an electron acquiring its rest mass energy equivalent over the distance of one reduced Compton wavelength $\lambdabar_C$. The critical QED field and corresponding acceleration are
\begin{equation}\label{eq:QEDcrit}
E_\mathrm{QED} = \frac{mc^2}{e\lambdabar_C} = \frac{m^2c^3}{e\hbar}\,,\qquad a_\mathrm{QED} =\frac{eE_\mathrm{QED}}{m}= \frac{mc^3}{\hbar} \, .
\end{equation}
Comparing with $a_\mathrm{RR}$ \req{aclas} we find
\begin{equation}\label{limitratio}
\frac{a_{\text{RR}}}{a_\mathrm{QED}}=\frac{3}{2} \alpha^{-1} = 205.5\,,\quad \alpha=\frac{e^2}{4\pi\varepsilon_0 \hbar c}\approx \frac{1}{137.0}\,.
\end{equation}
The classical critical field strength $E_\mathrm{CED}$ can be obtained by replacing the reduced Compton wavelength in \req{eq:QEDcrit} by the classical electron radius $r_0$,
\begin{equation}\label{eq:CEDcrit}
E_\mathrm{CED} = \frac{mc^2}{er_0} = E_\mathrm{QED} / \alpha
\end{equation}
and the relationship between the three scales for acceleration reads
\begin{equation}
a_\mathrm{RR} = \frac{3}{2} a_\mathrm{QED}/\alpha = \frac{3}{2}a_\mathrm{CED} \, .
\end{equation}
Therefore, apart from the numerical factor $3/2$ the EFO limiting acceleration is equivalent to the acceleration corresponding to the classical critical field scale.

$E_\mathrm{QED}$ arises in quantum electrodynamics as the benchmark field strength characterizing the instability of a strong electromagnetic field capable of rapid spontaneous decay into electron-positron pairs. Such quantum particulization of field energy has no classical analog. This field ``decay'' into particle pairs appears on first sight to lead to an effective quantum upper limit of acceleration 200 times smaller when compared to the here presented limiting value $a_{\text{RR}}$.

First we recall that a detailed exploration of experimental conditions carried out over past decades shows that experimental detection of vacuum field decay into pairs remains to this day exceedingly difficult. Present-day particulization experiments appear to struggle generating even one particle pair, let alone many pairs capable of back-reacting and ``neutralizing'' the applied field. Even so, the effort to experimentally identify field decay into pairs continues~\cite{Popov:2020xmd}.

Moreover, field decay into particle pairs does not occur for all field configurations and hence this quantum acceleration limit is not universal. The key exceptions to the applicability of the quantum limit are as follows:
\begin{enumerate}
\item 
Quantum instability does not afflict strong magnetic-field-dominated environments~\cite{Heisenberg:1936}, i.e. when $\mathcal{ B}^2>\mathcal{E}^2$. Moreover, the presence of an anomalous magnetic moment suppresses this instability~\cite{Evans:2018kor} for $\mathcal{B}^2<\mathcal{E}^2$.
\item 
Schwinger~\cite{Schwinger:1951nm} has shown that monochromatic plane waves of arbitrary strength and wavelength pass through the vacuum without pair production. The arguments Schwinger presented are based in the nature of light-wave and space-time symmetry, and very likely remain valid at any level of nonperturbative QED. Another way to recognize this is to note that both field invariants \reqs{eq:S}{eq:P} characterizing the plane wave are zero. However, charged particles riding such waves experience, according to the Lorentz force, an acceleration of arbitrarily large strength. 
\item 
Another ultra-strong acceleration occurring without pair production arises in the study of relativistic nuclear (heavy ion) collisions. In order to create particle-antiparticle pairs, the fields of the nuclei must exist for a long enough time~\cite{Greiner:1985ce}. Moreover, the potential well which is associated with any transitory field configuration must be capable of supplying to each produced pair the energy required for pair materialization. 
\end{enumerate}

We thus believe that the computation of classical radiation emission by charged particles we presented can be a physically meaningful model, establishing a classical RR force limit over two-hundred times greater than the quantum particulization limit,~\req{limitratio}. Even so, several further questions emerge when considering quantum dynamics:
\begin{enumerate}
\item
Is the classical method of establishing radiation limit permissible or will it be in a decisive way superseded by the inclusion of quantum emission effects in the radiation reaction?

\item
Since field configurations and kinematic conditions exists that allow us to bypass the well-known quantum limit to field strength, we should ask if indeed a universal limiting acceleration can be a feature of a complete theory. If it is as our discussion suggests, then very likely the quantum theory also will require modification to allow for such extreme strengths of applied force.

\end{enumerate}
 
As this work clarified, to answer these questions we must push the laboratory experimental conditions into the domain where the classical radiation limit becomes with certainty inapplicable and the actual experimental outcome will differ decisively from our classical model. Given that particles experiencing large acceleration usually have a large laboratory energy, they often satisfy, to a good approximation, classical dynamics. Therefore special effort needs to be made to find a parameter niche where quantum dynamics dominates. 

In general, a particle is capable of entering the quantum regime when the applied forces in the comoving frame surpass the quantum critical field strength over the space-time domain larger than the Compton wavelength of the particle. In this regime a new ``critical acceleration'' test of QED could be considered. In preparation for this step, in this work we have explored classical particle dynamics providing a reference for a parallel study in the context of QED. Our work clarifies how RR can self-consistently constrain the strength of applied forces. 

For strong fields and accelerations above the critical value, quantum corrections to the radiation reaction force are known to be important~\cite{Dipiazza2012}. To first order in $\tau_0$, the LL and EFO equations are known to be identical as we have also demonstrated here. Both of these equations are consistent with the classical limit of leading order QED calculations as shown in~\cite{Ilderton:2013}. However, we still view the question of connecting classical and quantum radiation reaction dynamics for strong accelerations as an open one.

\section{Conclusions}\label{Sec:Final}
The understanding of interactions between charged particles and strong electromagnetic fields is a problem of fundamental importance, and one that has occupied physicists for over a century. Particles in strong fields will experience a strong acceleration and emit radiation; the backreaction of this radiation on the particle can then play a significant role in the dynamics.

In this work we have demonstrated that, for nearly all cases considered, an upper limit to the acceleration of classical charged particles can be introduced through the Eliezer-Ford-O'Connell radiation reaction force. We have obtained an analytical formula for the invariant acceleration magnitude for both a particle in a constant, homogeneous electromagnetic field~\req{eq:a2resummed}, as well as in a plane wave field~\req{eq:a2_wave}. A notable exception to the limiting acceleration is the case of a particle uniformly accelerated by an electrical field (1D hyperbolic motion), which does not experience RR effects according to the EFO equation. Both the LAD and LL equations predict the same behavior for this case. 

The EFO limiting acceleration~\req{aclas} arises in the strong field, large velocity limit and has a value of $c/\tau_0$ leading via the Larmor formula to a limiting rate of radiation emission of $mc^2/\tau_0$. We have compared solutions of the EFO equation with the LL equation for a charged particle subjected to several different external forces, showing that the two theories are equivalent for weak accelerations. One can expand the EFO equation of motion to lowest order in $\tau_0$ to obtain the LL dynamics. For an example of this, see~\req{eq:amu_ours} and~\req{eq:a_LL_medium}. In this sense, the EFO equation will yield the same solutions as the LL equation within the domain where the LL approximation can be applied. The EFO equation is therefore equally valid as the LL equation as a perturbative model of the RR force. 

In the strong acceleration domain, the LL approximation itself breaks down as seen explicitly in the example of 1D material friction force in Sec.~\ref{sec:material_friction}. The EFO equation, however, yields more palatable limiting acceleration solutions. We can then view the EFO equation as a superior alternative to the LL equation in the strong acceleration domain.

Irrespective of the above considerations, there remains in any current RR force formulation the problem seen at the end of Sec.~\ref{sec:EConst}, where we have recognized that 1D hyperbolic motion accompanied by radiation emission has a vanishing RR EFO force. This is a universal defect of the current theoretical RR force picture seen in LAD, LL, and EFO equations of motion. In our opinion, this behavior should be absent in a completely self-consistent theory of charged particle dynamics. 

An upper bound on acceleration is not a universal feature of the EFO equation for all field configurations. However, the RR force is a phenomenological add-on to the Lorentz force, and the EFO form may not be accurate for all field configurations. In seeking a more complete description of RR theory we could postulate a limiting acceleration as a fundamental characteristic of charged particle motion and not an occasional feature, as it appears in the EFO model. In the future, we will return to this problem by modifying the EFO equation with the aim of developing a fully consistent description of radiation emission and classical particle dynamics in which limiting acceleration is taken as a fundamental principle. We see two possible approaches toward this end:
\begin{enumerate}
\item
The limiting acceleration feature of the EFO equation is reminiscent of the Born-Infeld limiting field theory and one may explore how a limiting field strength can, in some circumstances, lead to a limiting force which imposes a limit on the acceleration. Moreover, the question of radiation reaction in Born-Infeld theory is an intriguing one since the limiting field regularizes the divergence usually present in the self-force~\cite{Kiessling:2019eip}. It remains to be seen if consistency can be established between the EFO radiation reaction model and a limiting EM field strength model. 
\item
The path warping method developed in~\cite{Formanek:2020zwc} allows us to relax the four-dimensional orthogonality constraints on covariant equations of motion. This offers an additional freedom in formulating a RR equation of motion, allowing the introduction of limiting acceleration as a guiding principle to formulate a RR force. We hope that this will assist in the development of a RR model in which limiting acceleration appears for all cases of applied force.
\end{enumerate}

The path warping method is a step out of the set of conventional RR theories discussed in this work. However, it is an additional promising method that can be combined with either the EFO equation or extended into the quantum domain to create a more consistent RR formulation with limiting acceleration. In Ref.\,\cite{Formanek:2020zwc} path warping for particle motion in a medium was presented as a dynamical deformation of the medium induced by motion of the particle. Such a deformation absorbs kinetic energy which is ultimately released as radiation. While special relativity forbids ``\ae ther warping'' by velocity, a similar mechanism can operate driven by acceleration or force allowing the particle path to be impeded by radiation energy loss for any form of applied force.

We have also discussed in Sec.~\ref{sec:quantum} the 200 times lower limit to acceleration introduced by strong field ``decay'' into particle pairs. The associated ``quantum'' limiting field strength introduces an effective quantum upper limit to acceleration which does not occur in all field configurations of experimental interest. It remains to be seen how a quantum theory of radiation reaction will change the classical results presented here and if a quantum limit to acceleration can be introduced directly through radiation emission rather than indirectly through pair production.

In conclusion, we believe that the EFO equation is, among all classical descriptions of RR, the most promising. Taking seriously the new insights about RR presented here, and in particular the upper limit to acceleration, we believe that exploration of a quantum theory of radiation reaction built upon a classical theory with limiting acceleration would be an appropriate fresh start. 

\acknowledgements We thank the referee for helpful and insightful comments and suggestions.


\end{document}